\documentclass[superscriptaddress,aps,prl,reprint,twocolumn,amsmath,amssymb]{revtex4-2}
\usepackage{graphicx}
\usepackage{hyperref}
\usepackage{amssymb}
\usepackage{slashed}
\usepackage{dcolumn}
\usepackage{amsmath}
\usepackage{bm}
\usepackage{colordvi}
\usepackage{algorithm}
\usepackage{algpseudocode}
\usepackage{multirow}
\usepackage{titlesec}
\usepackage{newtxtext,newtxmath}
\usepackage{dsfont}
\usepackage{xcolor}
\usepackage{mathbbol}

\usepackage{hyperref}
\hypersetup{
    colorlinks=true,
    linkcolor=blue,
    citecolor=blue,     
    urlcolor=blue,
}

\allowdisplaybreaks

\usepackage{mathrsfs}
\makeatletter

\newcommand{\Rmnum}[1]{\expandafter\@slowromancap\romannumeral #1@}
\makeatother

\begin{document}

\title{Microscopic phase-transition theory of charge density waves: revealing hidden {crossovers} of phason and amplitudon}

\author{F. Yang}
\email{fzy5099@psu.edu}

\affiliation{Department of Materials Science and Engineering and Materials Research Institute, The Pennsylvania State University, University Park, PA 16802, USA}

\author{L. Q. Chen}
\email{lqc3@psu.edu}

\affiliation{Department of Materials Science and Engineering and Materials Research Institute, The Pennsylvania State University, University Park, PA 16802, USA}

\date{\today}

\begin{abstract} 
  We develop a self-consistent phase-transition theory of charge density waves (CDWs), starting from a purely microscopic model. Specifically, we derive a microscopic CDW gap equation $|\Delta_0(T)|$, taking into account of thermal phase fluctuations (i.e., thermal excitation of phason) and their influence on CDW pinning (i.e., the phason mass) and CDW gap. We demonstrate that as temperature increases from zero, the phason gradually softens, leading to a thermal depinning {crossover} (where the phason becomes gapless) at $T_d$ and a subsequent first-order CDW phase transition at $T_c>T_d$. The predicted values of $T_d$, $T_c$ as well as the large ratio of $|\Delta_0(T=0)|/(k_BT_c)$ for the quasi-1D CDW material (TaSe$_4$)$_2$I show quantitative agreements with experimental measurements and explain many of the previously observed key thermodynamic features and unresolved issues in literature. To further validate the theory, we calculate the energy gap of CDW amplitudon and its lifetime, and reveal a {crossover} of amplitudon from a lightly damped to a heavily damped excitation during pinning-depinning {crossover}  while its energy gap is nearly unchanged throughout the entire CDW phase. This finding quantitatively captures and explains the recently observed coherent signal in ultrafast THz emission spectroscopy on (TaSe$_4$)$_2$I. 
  
\end{abstract}

\maketitle  

{\sl Introduction.---}Many-body interactions in quantum materials give rise to a variety of intriguing phenomena associated with rich electronic phases~\cite{coleman2015introduction}. One notable example is superconductivity, where electrons with opposite momenta and spins form Cooper pairs via a phonon-mediated effective attractive potential, as described by BCS superconductivity theory~\cite{bardeen1957theory,schrieffer1964theory}. The Cooper pair condensation below a critical temperature establishes the superconducting order, leading to the emergence of Bogoliubov quasiparticles~\cite{bardeen1957theory,schrieffer1964theory} and collective excitations (e.g., Higgs~\cite{littlewood1981gauge,pekker2015amplitude,matsunaga2013higgs,matsunaga2014light,yang2019gauge,shimano2020higgs,sun2020collective,yang2020theory} and Nambu-Goldstone modes~\cite{nambu1960quasi,goldstone1961field,goldstone1962broken,nambu2009nobel,yang2019gauge,ambegaokar1961electromagnetic,littlewood1981gauge,sun2020collective}). Then, the quasiparticle description within a self-consistent quantum statistical framework~\cite{abrikosov2012methods,schrieffer1964theory} enables predictions/calculations of the finite-temperature properties of superconductors (e.g., temperature dependence of the superconducting gap) and critical behavior of the superconducting phase (e.g., transition temperature), based solely on the knowledge of the BCS ground state.

Another representative example of electronic condensation is the formation of charge density waves (CDWs)~\cite{gruner1985charge,gruner1988dynamics,lee1974conductivity}, which is triggered by the electron-phonon interaction to stabilize electronic and/or
structural instabilities, e.g., Peierls instability~\cite{peierls1996quantum}. This condensation leads to a periodic modulation of the electronic charge density accompanied by a periodic lattice distortion. The formation of CDW induces a gap opening in the electron energy spectrum at the Fermi level, driving a metal–insulator transition~\cite{gruner1985charge,gruner1988dynamics,lee1974conductivity}. Meanwhile, two types of collective excitations emerge below the CDW transition temperature~\cite{gruner1985charge,gruner1988dynamics,lee1974conductivity}, corresponding to the amplitude and phase fluctuations of the complex CDW order parameter~\cite{pekker2015amplitude}, as illustrated in Fig.~\ref{figyc}a.  The gapped amplitude mode, called amplitudon, resembles the Higgs boson in field theory, while the gapless phase mode, called phason, corresponds to the Nambu-Goldstone bosons due to  spontaneous breaking of continuous symmetry~\cite{nambu1960quasi,goldstone1961field,goldstone1962broken,nambu2009nobel}. The presence of the gapless phase mode suggests that an applied electric field can induce a current-carrying sliding motion of the CDW, as originally proposed by  Fr\"ohlich~\cite{frohlich1954theory}. However, impurities~\cite{fukuyama1978dynamics}, commensurability effect or lattice imperfections~\cite{gruner1985charge,gruner1988dynamics} can pin the CDW by introducing an excitation gap (i.e., effective mass) in the phase mode. {This excitation gap regularizes the infrared divergence of the phase correlations underlying the Mermin-Wagner theorem~\cite{hohenberg1967existence,mermin1966absence,coleman1973there}, which forbids long-range order in low-dimensional systems, and consequently,  the phase-correlation function becomes short-ranged.}  As a result, to depin the CDW from defects and enable its sliding motion, the applied electric field must exceed a threshold value to overcome the pinning energy~\cite{gruner1988dynamics}, leading to a nonlinear DC conductivity, as widely observed in experiments~\cite{maki1983charge,liu2018low,li2023influence}.
  Recent advances in ultrafast pump-probe techniques have opened the possibility and growing strong interest of coherently driving the collective modes (amplitudon and phason) of CDWs in the nonlinear regime~\cite{torchinsky2013fluctuating,kim2023observation}, offering a new pathway to manipulate the CDW phase using intense laser fields~\cite{yoshikawa2021ultrafast}.

{Despite significant and continuously growing experimental progress, theoretical frameworks capable of adequately describing CDW formation, quantitatively capturing their critical behavior, and  self-consistently accounting for the dynamics of collective excitations and their responses to external stimuli remain limited. In practice, material-specific computations  have relied predominantly on first-principles density-functional theory (DFT) calculations}~\cite{joshi2019short,christensen2021theory,wang2022origin}. While DFT has proven highly effective and valuable in analyzing the electronic and structural stability of the ground state, it faces limitations in providing accurate and consistent predictions of properties across the entire temperature range, capturing critical behaviors and describing the nonequilibrium dynamics of excited quasiparticles and collective excitations. Addressing these challenges requires a quantum statistical description that builds upon the DFT-determined or the measured ground-state properties.

An early study by Lee, Rice, and Anderson proposed a ground state of CDWs formally analogous to the BCS ground state of superconductors~\cite{lee1974conductivity} and derived the energy spectra of amplitudon and phason at zero temperature using Green function approach. However, despite this foundation, a microscopic phase-transition theory for CDWs remains elusive. One key challenge is that the mean-field  phase-transition theory inherently predicts a second-order phase transition, as observed in conventional superconductors~\cite{bardeen1957theory,schrieffer1964theory}, whereas many of CDW materials exhibit a first-order transition~\cite{monceau2012electronic}. Another limitation concerns the ratio of the zero-temperature CDW gap to the transition temperature, $2|\Delta_0(T=0)|/k_BT_c$. The mean-field theory predicts a value of $3.52$~\cite{bardeen1957theory,schrieffer1964theory}, significantly smaller than experimentally measured ones in CDWs~\cite{sinchenko2012sliding}, e.g., 11.2 for NbSe$_3$~\cite{orlov2006interaction,monceau2012electronic}; 17.68 for (TaSe$_4$)$_2$I~\cite{huang2021absence,monceau2012electronic,gooth2019axionic}; 13-14 for (NbSe$_4$)$_{3}$I, 13.7 for (NbSe$_4$)$_{10}$I$_3$~\cite{monceau2012electronic}; 13.95 for TaS$_3$~\cite{brill1982elastic,monceau2012electronic}.

In fact, early  studies~\cite{lee1973fluctuation,rice1981impurity,maki1986thermal} proposed that, as the temperature increases from zero, the CDW undergoes a depinning {crossover} from a pinned to a sliding state at a critical temperature $T_d$ below $T_c$, corresponding to the softening of the phason into a gapless excitation.  In this scenario, the sliding CDW is expected to exhibit similarities to the so-called phase-fluctuating superconductivity~\cite{yang2021theory,emery1995importance,benfatto2001phase,pracht2016enhanced,fisher1990presence,fisher1990quantum,curty2003thermodynamics,li2021superconductor,dubi2007nature,sacepe2020quantum,sacepe2008disorder} in low-dimensional superconductors, extending beyond the mean-field paradigm.

Inspired by this insight and recently developed microscopic phase-transition theory of phase-fluctuating superconductivity~\cite{yang2021theory}, here we develop a self-consistent phase-transition theory of CDWs by treating the CDW gap and phase on an equal footing within the path-integral framework, starting from a purely microscopic model. This theory incorporates the microscopic CDW gap equation and thermal phase fluctuations (i.e., thermal excitation of phason), enabling calculations of finite-temperature CDW properties and, in particular, critical behaviors such as the depinning {crossover} and CDW phase transition, based solely on the knowledge of the ground-state properties determined by experiments or DFT calculations. As a specific application, we consider the quasi-one-dimensional (quasi-1D) CDW material (TaSe$_4$)$_2$I. We show that as temperature increases from zero, the phason gradually softens, resulting in a depinning {crossover} (gapless phason) at $T_d$. With further temperature increase, significant thermal phase fluctuations suppress the CDW gap, ultimately driving a first-order phase transition at $T_c$, well below the mean-field transition point. This first-order transition results in a significantly enhanced ratio $2|\Delta_{0}(T=0)|/k_BT_c$ compared to the BCS prediction. Notably, for (TaSe$_4$)$_2$I, our theory predicts a depinning {crossover} at $T_d\approx160~$K,
a first-order CDW phase transition at $T_c\approx268~$K, and a large $2|\Delta_{0}(T=0)|/k_BT_c\approx17.3$, in remarkably quantitative agreement with experimental measurements. These findings also provide a comprehensive explanation for many of the previously observed key thermodynamic features.

To further validate our theory, we derive the energy spectrum of the amplitudon and show that its interaction with the phason induces a damping. As a result, at the depinning {crossover} ($T_d$),  where the phason fully softens into a gapless excitation with a significantly increased phason number, the amplitudon {crosses  over} from a lightly damped to a heavily damped excitation, while its excitation gap remains nearly unchanged throughout the entire CDW phase. This finding quantitatively explains the recently observed nonlinear coherent signal in THz-emission spectroscopy of  (TaSe$_4$)$_2$I~\cite{kim2023observation}, which exhibits a temperature-independent coherent frequency but a strongly temperature-dependent signal strength, gradually diminishing as $T$ approaches $T_d$ and vanishing above $T_d$.   

{\sl Theoretical model.---}We begin with the Fr\"ohlich's Hamiltonian~\cite{frohlich1954theory} and single out the critical phonon mode at $Q=2k_F$ with the energy $\omega_{Q}$, as studied by Lee, Rice and Anderson~\cite{lee1974conductivity}. The lattice distortion is then described by the order parameter 
{\begin{equation}
\Delta=2g\langle{b_{Q}}\rangle,
\end{equation} }
where $b_{Q}$ is the phonon annihilation operator and $g$ denotes the electron-phonon interaction strength. This {{lattice-distortion-driven}} order parameter is generally complex: $\Delta=|\Delta|e^{i\theta}$, where $|\Delta|=|\Delta_0|+\delta|\Delta|$ and $\theta=\theta_0+\delta\theta(R)$. Here, $|\Delta_0|$ and $\theta_0$ correspond to the CDW gap and phase, respectively, while $\delta|\Delta|$ and $\delta\theta(R)$ represent the amplitude and phase fluctuations. The thermal average of phase fluctuations satisfies $\langle\delta\theta\rangle=0$. {{Through the electron–phonon interaction, the lattice  distortion opens a gap in the electronic spectrum and produces a periodic modulation of the electronic charge density}~\cite{gruner1985charge,gruner1988dynamics,lee1974conductivity}:
\begin{equation}
\rho(R)=\rho_0+|\Delta_0|\rho_0\cos(QR+\theta_0)/(\lambda\varepsilon_{k_F}).  
\end{equation}
To explicitly treat the thermodynamics of the CDW gap $|\Delta_0(T)|$ and phase fluctuations $\delta\theta$, a rigorous derivation within the path-integral approach is performed (see Supplemental Materials, including Refs.~\cite{yang2023optical,yang2018gauge,yang2024optical,Landaubook,peskin2018introduction,fujishita1986neutron,lin2024unconventional,yi2021surface,tournier2013electronic,larkin1965zh,NASRETDINOVA2015180,nasretdinova2009electric,PhysRevB.48.1368,PhysRevB.108.045148,brazovskii1975phase}). The electronic sector enters  through the electron-phonon interactions, influencing 
lattice-distortion (phonon-condensation) process and hence gap equation. We summarize the derived self-consistent microscopic phase-transition theory, consisting of the CDW gap equation:
\begin{equation}\label{gap}
\frac{1}{\lambda}=-\int^{\omega_D}_{-\omega_D}\frac{f\big(E_k\!+\!v_Fp_s\big)\!-\!f\big(-E_k\!+\!v_Fp_s\big)}{2E_k}{d\varepsilon_k},  
\end{equation}
and thermal phase fluctuations of $p_s=\partial_R\delta\theta(R)/2$ given by 
\begin{equation}\label{tpf}
\langle{p_s^2}\rangle=\int\frac{dq}{2\pi}\frac{q^2n_B\big(\Omega_{{\rm P}}(q)\big)}{D_0^*\Omega_{{\rm P}}(q)}.  
\end{equation}
Here, $\lambda=2D_0g^2/\omega_Q$ is the dimensionless coupling constant~\cite{gruner1985charge,gruner1988dynamics} and $D_0^*=D_0m^*/m$ is the renormalized density of states,  with $D_0=1/(\hbar{\pi}v_F)$ being the bare density of states at Fermi level; $f(x)$ and $n_B(x)$ are Fermi-Dirac and Bose-Einstein distributions; the energy spectrum of fermionic quasiparticles {\small $E_k=\sqrt{\varepsilon_k^2+|\Delta_0|^2}$}  with $\varepsilon_k=\frac{\hbar^2(k_F+k)^2}{2m}-E_F ={\hbar}v_Fk$; the energy spectrum of the bosonic phason $\Omega_{\rm P}(q)$ is given by 
\begin{equation}\label{phasonenergy}
\Omega_{\rm P}^2(q)=\frac{m}{m^*}f_sm^2_{\rm P}(T)+\frac{m}{m^*}{f_s{\hbar}^2q^2v_F^2}, 
\end{equation}
where $f_s$ is the condensation fraction~\cite{hayashi1996topological,hayashi2000ginzburg} written as 
\begin{equation}
  f_s=\int{d\varepsilon_k}\frac{|\Delta_0|^2}{E_k}\partial_{E_k}\big[\frac{f(E_k\!+\!v_Fp_s)\!-\!f(v_Fp_s\!-\!E_k)}{2E_k}\big],
\end{equation}
with $f_s=1$ near $T=0$ and $f_s\propto{|\Delta_0|^2}/{(\pi{T})^2}$ for $|\Delta_0|\rightarrow0$. ${m^*}/{m}=1+{4|\Delta_0|^2}/({\lambda\omega^2_Q})$, the same as the one derived in previous works~\cite{gruner1985charge,gruner1988dynamics,lee1974conductivity}; $m_{\rm P}(T)$ represents the excitation gap of phason, i.e., the phason mass, induced by CDW pinning due to impurities~\cite{fukuyama1978dynamics}, a commensurability effect or lattice imperfections~\cite{gruner1985charge,gruner1988dynamics}. To account for the effect of thermal phase fluctuations on CDW pining, we employ the self-consistent-field approximation, following the early works~\cite{maki1986thermal,rice1981impurity} by assuming that the phason mass follows a self-consistent temperature dependence $m^2_{\rm P}(T)=m^2_{\rm P}(T=0)\exp(-{\xi^2}p_s^2)$, where $\xi(T)\propto\hbar{v_F}/|\Delta_0(T)|$ is the coherence length. This dependence  describes the phason softening  as temperature increases.

The key point of the theory is the vanishing thermal average $\langle{p_s}\rangle$ but nonzero $\langle{p_s^2}\rangle$, and the CDW gap equation in Eq.~(\ref{gap}) is an even function of $p_s$. As a result, the theory addresses the influence of thermal phase fluctuations on CDW gap by introducing the Doppler shift $v_Fp_s$~\cite{fulde1964superconductivity,yang2018fulde,yang2021theory,landau1941theory} and on CDW pining through the phason mass $m_{\rm P}(T)$ in a self-consistent manner. As the temperature rises from zero, thermal phase fluctuations gradually emerge due to the bosonic thermal excitation of the gapped (massive) phason, which weakens the CDW pinning (softens the phason). When the temperature increases above a certain temperature scale, 
the phason mass becomes minimal {(exponentially small)}, signaling the {crossover} into a {nearly} gapless excitation and thus marking the CDW depinning {crossover} from a pinned to a sliding state. As temperature further increases, phase fluctuations continue to grow and suppress CDW gap via Doppler shift, ultimately driving a CDW phase transition at critical temperature $T_c$.

{\sl Results.---}We next focus on a classical quasi-1D CDW material, (TaSe$_4$)$_2$I, to perform numerical simulation. The tunneling spectroscopic measurements of (TaSe$_4$)$_2$I reported a CDW gap of $\sim200~$meV~\cite{huang2021absence,gooth2019axionic,forro1987hall,wang1983charge}. Other zero-temperature parameters used in our simulations, along with their determination based on several independent measurements, are provided in the Supplemental Materials~\cite{supple}.  The numerically predicted temperature dependencies of the CDW gap $|\Delta_0(T)|$ and phason mass $m_{\rm P}(T)$ are plotted in Fig.~\ref{figyc}b. With an increase in temperature from zero and hence increased thermal phase fluctuations from zero (inset of Fig.~\ref{figyc}b), the phason mass $m_{\rm P}(T)$ gradually softens and {nearly} vanishes at temperatures above $T_d\approx160~$K, marking the depinning {crossover} (gapless phason), while the CDW gap remains nearly unchanged for $T<T_d$. Above $T_d$, significant thermal phase fluctuations begin to suppress the CDW gap, resulting in a noticeable deviation from the mean-field behavior (which is driven by the thermal excitation of fermionic quasiparticles, shown by the dotted curve in the inset of Fig.\ref{figyc}b). Notably, as temperature further increases, thermal phase fluctuations destroy the CDW gap and drive a first-order phase transition when the fluctuation strength $|v_Fp_s(T)|$ exceeds the order-parameter strength $|\Delta_0(T)|$ at $T_c$ (as shown in the inset of Fig.\ref{figyc}b), which is considerably lower than the expected mean-field transition temperature.  Several early  experiments~\cite{Lorenzo_1998,huang2021absence,suzuki1988effects} have reported a first-order CDW phase transition in (TaSe$_4$)$_2$I. This behavior actually aligns with the fluctuation-driven transitions in low-dimensional systems~\cite{PhysRevB.87.134407,PhysRevLett.69.1085,PhysRevB.107.075130,PhysRevB.96.094419}, where fluctuation effects are essential and can modify the transition away from the mean-field second-order expectation, in some cases yielding the weakly first-order characteristics. For 1D CDWs, the nearly gapless phason due to softening reactivates the infrared-divergent nature of phase fluctuations and drives the deviation from mean-field behavior.

\begin{widetext}
  \begin{center}
\begin{figure}[htb]
  {\includegraphics[width=16.5cm]{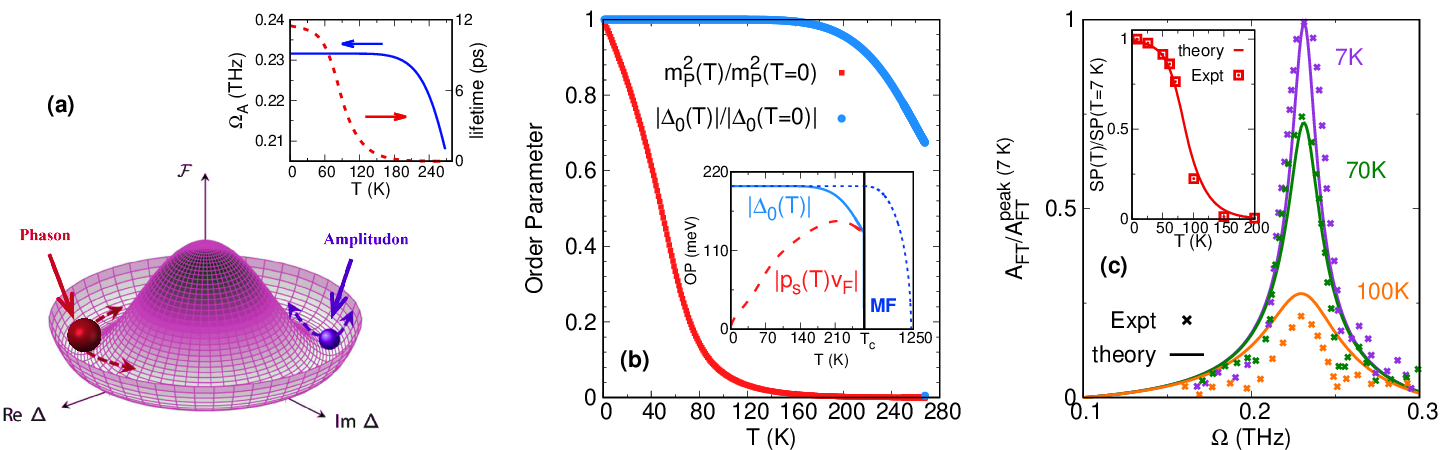}}
  \caption{{{({\bf a}) Schematic illustration of the amplitudon and phason in CDWs by the Mexican-hat-potential free energy. The inset shows the calculated excitation gap and lifetime of amplitudon in (TaSe$_4$)$_2$I. ({\bf b}) Calculated CDW gap and phason mass of (TaSe$_4$)$_2$I as a function of temperature. The inset shows the CDW gap $|\Delta(T)|$ (solid curve) and Doppler shift $|p_s(T)v_F|$ (dashed curve) from our theory, and, the mean-field-theory gap (dotted curve), as a function temperature. ({\bf c}) Theoretically predicted (curves) and experimentally measured (crosses) coherent signal in THz emission spectroscopy of (TaSe$_4$)$_2$I at different temperatures, normalized to the peak value at 7~K. The inset shows the theoretical (curve) and experimental (squares) spectral weight of the coherent signal versus $T$. Experimental data come from Ref.~\cite{kim2023observation}.  }} }    
\label{figyc}
\end{figure}
\end{center}
\end{widetext}

Exactly due to the occurrence of the first-order phase transition here, the ratio of the zero-temperature CDW gap to the CDW phase transition temperature, $2|\Delta_0(T=0)|/k_BT_c$, significantly exceeds the mean-field prediction of 3.52. Notably, for  (TaSe$_4$)$_2$I, using only the ground-state (zero-temperature) parameters without any other free parameters, our theory predicts a depinning {crossover} at $T_d\approx160~$K, a first-order CDW phase transition at $T_c=268~$K and a large ratio $2|\Delta_0(T=0)|/k_BT_c=17.36$, and these predictions are in excellent agreement with experimentally measured values: $T_d\sim160~$K~\cite{li2023influence}, $T_c\approx263~$K~\cite{huang2021absence,monceau2012electronic,gooth2019axionic,brutting1995dc,Lorenzo_1998,PhysRevLett.132.266504} and $2|\Delta_0(T=0)|/k_BT_c=17.68$~\cite{huang2021absence,monceau2012electronic,gooth2019axionic}. The predicted depinning {crossover} at $T_d=160~$K, marked by the emergence of {nearly} massless phason excitations, also provides a natural explanation for the experimentally observed {nearly} vanishing electric-field threshold for dc conductivity above 160~K~\cite{li2023influence}.

To further validate our theory, we derive the energy spectrum $\Omega_A$ of amplitudon (i.e., amplitude mode $\delta|\Delta|(t)$ of the CDW order parameter, as illustrated in Fig.~\ref{figyc}a), determined by
\begin{equation}\label{amplitudon}
\Omega^2_{\rm A}=\frac{{\lambda\omega^2_Qf_s}}{1+\lambda\omega^2_Qf_s/(4|\Delta_0|^2)}.  
\end{equation}
At zero temperature with $f_s=1$ and $2|\Delta_0|\gg{\sqrt{\lambda}\omega_Q}$, the derived $\Omega_{\rm A}$ reduces to $\sqrt{\lambda}\omega_Q$, the same as the one obtained in previous works at $T=0$~\cite{gruner1985charge,gruner1988dynamics,lee1974conductivity}. Conversely, if one assumes an artificially small $|\Delta_0|$, the derived $\Omega_{\rm A}$ approximates $2|\Delta_0|$, consistent with the predictions from the phenomenological Landau phase-transition theory~\cite{littlewood1981gauge,pekker2015amplitude} and analogous to the Higgs/amplitude mode (with excitation energy $\omega_H=2|\Delta_0|$) in superconductors~~\cite{littlewood1981gauge,pekker2015amplitude,matsunaga2013higgs,matsunaga2014light,yang2019gauge,shimano2020higgs,sun2020collective,yang2020theory}. The temperature dependence of $\Omega_{\rm A}$ is plotted in the inset of Fig.~\ref{figyc}a. As expected, $\Omega_{\rm A}$ closely follows the behavior of the CDW gap. It remains nearly unchanged at $0.231~$THz below $T_d=160~$K, and gradually decreases above $T_d$, reaching $0.208~$THz at $T_c$.

Unlike in superconductors, where the amplitude and phase modes are decoupled due to particle-hole symmetry~\cite{pekker2015amplitude}, the amplitudon and phason in CDWs are intrinsically coupled (see Supplemental  Materials). This coupling introduces a damping term in the equation of motion of the amplitudon $\delta|\Delta(t)|$, and this equation of motion in the frequency space ({\small $t\rightarrow\Omega$}) reads
\begin{equation}\label{EoA}
(\Omega^2-i\Omega\gamma-\Omega^2_{\rm A})\delta|\Delta(\Omega)|=S_e.
\end{equation}
Here, $S_e$ represents the source term by an external stimulus; $\gamma$ denotes the damping rate of amplitudon resulting from its coupling with phason. While a rigorous microscopic calculation~\cite{yang2024diamagnetic,yang2020influence,cui2019impact} of this damping is complex and beyond the scope of our study, within the
microscopic mechanism for calculating the scattering probabilities of boson-emission and -absorption processes of the scattered particles~\cite{yang2015hole,yang2016spin,yang2025THz}, we approximate the damping rate as $\gamma(T)=\gamma_{th}(T)+\gamma_0$, where $\gamma_0$ is a temperature-independent constant, and $\gamma_{th}(T)=c_0\int\frac{dq}{2\pi}\frac{n_B[\Omega_{{\rm P}}(q)]}{D_0^*\Omega_{{\rm P}}(q)}$ accounts for contributions from the thermal excitation of the phason, i.e., phason number. Taking $1/\gamma_0\approx11.47~$ps and $1/c_0\approx1.02~$ps, we plot the temperature dependence of the amplitudon lifetime in the inset of Fig.~\ref{figyc}a.  As temperature rises from zero to $T_d=160~$K, the softening of the phason (i.e., its {crossover} from a gapped to a gapless excitation) causes $1/\gamma(T)$ to rapidly decrease from a finite value to nearly zero. This indicates that the amplitudon {crosses over} from a lightly damped to a heavily damped excitation.

Recently, optical pump-probe measurements on (TaSe$_4$)$_2$I have reported an excited coherent signal in THz emission spectroscopy~\cite{kim2023observation}. This signal exhibits a temperature-independent coherent frequency, while its strength gradually diminishes as the temperature approaches
$T_d$ from zero and vanishes entirely above $T_d$. Based on this behavior, we deduce that the observed signal originates from the coherent excitation of the amplitudon. Using Eq.~(\ref{EoA}), considering short-pulse external stimulus (i.e., constant $S_e$) and assuming the experimentally observed signal is proportional to $\delta|\Delta(\Omega)|$, we compare the theoretical predictions with experimental data~\cite{kim2023observation} in Fig.\ref{figyc}c and its inset. As shown, the predicted frequency and temperature dependencies of the optically excited THz-emission coherent
signal from our theory align remarkably well with the experimental measurements below 70 K, and shows
certain small deviations at higher temperatures. Notably, this quantitative agreement is achieved only using zero-temperature parameters that were independently determined from experiments in low-temperature limits, without any temperature-dependent fitting.

{\it Discussion.}---The theory here 
closely resembles the phase-fluctuating superconductivity theory~\cite{yang2021theory,yang2024arXiv,xy6z-hxcv}, owing to their physical similarity~\cite{gruner1985charge,gruner1988dynamics,lee1974conductivity}. {The central idea is that the key physical ingredients (including the CDW gap, fermionic quasiparticles, phase fluctuations, and disorder-induced phase pinning) and their interplay give rise to a unique competition between gap stabilization, phason softening, and pinning-depinning behavior. This competition governs both the thermodynamics (e.g.,  the temperature evolution of the phase and the phase transitions) and the dynamics of collective excitations (phason and amplitudon).}  At $T=0$, the derived energy spectra of phason and amplitudon as well as gap equation exactly recover the ones derived by Lee, Rice and Anderson~\cite{gruner1985charge,gruner1988dynamics,lee1974conductivity}. Our theory enables calculations of
finite-temperature properties and, in particular, critical behaviors such as the depinning {crossover}  and CDW phase transition, based solely on the knowledge of ground-state properties. It also provides a route to derive the dynamics of collective excitations (amplitudon and phason) at finite temperatures and calculate ultrafast THz-optical responses  without additional phenomenology.

The calculation identifies a pinning-depinning {crossover} within the CDW phase below $T_c$ and in particular, reveals the consequences it triggers. It also resolves the long‐standing puzzle of why $2|\Delta(0)|/(k_B T_c)$ is anomalously large in CDW systems. While the phason mass arises from CDW pinning by impurities, commensurability effect or lattice imperfections, the depinning {crossover} temperature  $T_d$ is independent of the specific value of pinning strength $m_{\rm P}(T=0)$~\cite{maki1986thermal,rice1981impurity} and hence, is \emph{sample-independent}. This is because the depinning {crossover} satisfies the condition $\xi^2p_s^2(T=T_d)\gg1$, i.e., 
\begin{equation}
\xi^2\int\frac{dq}{2\pi}\frac{q^2n_B\big({\hbar}qv_F\sqrt{f_sm/m^*}\big)}{D_0^*{\hbar}qv_F\sqrt{f_sm/m^*}}\Big|_{T=T_d}\gg1,      
\end{equation}
which is clearly independent of $m_{\rm P}(T=0)$. In addition, within the self-consistent-field approximation~\cite{maki1986thermal,rice1981impurity}, the CDW depinning {crossover} described here corresponds to an {exponential crossover}. Consequently, no distinct thermodynamic signatures such as latent heat or specific-heat discontinuities emerge at $T_d$, consistent with existing experimental findings, making CDW depinning {crossover} a hidden {crossover}.  Nonetheless, the emergence of a massless phason above $T_d$ can manifest in thermal transport behaviors. For example, this effect explains the observed peak at 160~K in temperature dependence of Seebeck coefficient~\cite{surendranath1986temperature} and detected minimum at 160~K in thermal conductivity $\kappa$(110)~\cite{smontara2002anisotropy}.  Additionally, this massless phason can provide an additional relaxation channel for other excitations, resulting in a {crossover} from lightly damped to heavily damped behavior, as demonstrated by the amplitudon damping revealed in the present study.  

{The predicted features here due to phason softening, including the pinning–depinning crossover at $T_d$ (below $T_c$), the emergence of an additional relaxation channel due to the nearly massless phason above $T_d$, and the abrupt suppression of the CDW gap near $T_c$, arise from the fluctuation-feedback mechanism and represent generic features of lattice-driven CDWs. These features can, in principle, be detected by optical probes (e.g., pump–probe spectroscopy) that can directly access relaxation times and by spectroscopic probes such as ARPES and STM that can resolve the single-particle excitation gap.}

The experimental study in Ref.~\cite{kim2023observation} attributed the observed coherent signal in THz emission spectroscopy to the direct excitation of the phason, which becomes visible below $150~$K due to the phason acquiring a mass. 
 Since the coherent frequency (0.23~THz~\cite{Frequency}) corresponds to the mass of the excited mode, this interpretation is not likely, as it would require the nearly massless phason (above 150~K) to abruptly acquire a constant mass upon cooling below 150~K.  This scenario of a constant phason mass below 150~K also fails to explain why the signal amplitude increases upon cooling despite the coherent frequency remaining fixed.  Drawing on both previous experimental~\cite{matsunaga2013higgs,matsunaga2014light} and theoretical  studies~\cite{yang2019gauge,yang2020influence} on Higgs‐mode observations in superconductors, we find that the observed feature in the present CDW system is more naturally explained as the direct detection of the amplitudon, whose dynamics are nonetheless strongly affected by the softening of the phason. This coupling between the phason and amplitudon is a unique characteristic of CDW systems and has no counterpart in superconductors, where the Higgs and phase modes remain entirely decoupled.  {While the amplitudon is not dipole-active in an ideal, translationally invariant crystal, in real pinned CDWs (studied here), impurities, defects, lattice imperfections, grain boundaries and surface inevitably breaks translational symmetry. Such symmetry breaking can lead to a finite polarization associated with amplitude fluctuations, allowing the amplitudon to acquire optical activity and produce a measurable resonance at the amplitude-mode frequency~\cite{DONOVAN1990721,gruner1988dynamics}.}

{\it Acknowledgments.---}F.Y. and L.Q.C. acknowledge support from the US Department of Energy, Office of Science, Basic Energy Sciences, under Award Number DE-SC0020145 as part of Computational Materials Sciences Program. F.Y. and L.Q.C. also appreciate the  generous support from the Donald W. Hamer Foundation through a Hamer Professorship at Penn State.

%
\end{document}


\title{Microscopic phase-transition theory of charge density waves: revealing hidden {crossovers} of phason
and amplitudon (Supplementary Material)}

\author{F. Yang}
\email{fzy5099@psu.edu}

\affiliation{Department of Materials Science and Engineering and Materials Research Institute, The Pennsylvania State University, University Park, PA 16802, USA}

\author{L. Q. Chen}
\email{lqc3@psu.edu}

\affiliation{Department of Materials Science and Engineering and Materials Research Institute, The Pennsylvania State University, University Park, PA 16802, USA}

\maketitle 

\section{Microscopic model of 1D CDWs}

In this section, we present the microscopic model of the conventional 1D CDWs. We begin with the Fr\"ohlich Hamiltonian~\cite{frohlich1954theory,lee1974conductivity}:
\begin{equation}\label{Ham}
H=\sum_{k\sigma}\xi_{k}c^{\dagger}_{k\sigma}c_{k\sigma}+\sum_q{\omega_q}b_q^{\dagger}b_q+\sum_{qk\sigma}[gc^{\dagger}_{k+q\sigma}c_{k\sigma}(b_q+b_{-q}^{\dagger})+h.c.],
\end{equation}
where $c_{k\sigma}^{\dagger}$ and $b_q^{\dagger}$ are creation operators for a electron with spin $\sigma$ and momentum $k$, and for a phonon with momentum $q$, respectively; $c_{k\sigma}$ and $b_q$ are corresponding annihilation operators; $\xi_{k}=\frac{\hbar^2{k^2}}{2m}-\mu$ and $\omega_q$ are corresponding electron and phonon energies, respectively, with $\mu$ denoting the chemical potential; $g$ denotes the electron-phonon interaction strength. Based on this Hamiltonian, the action of the model is given by
\begin{equation}
  S=\int{dt}\Big\{\int{dx}\sum_{\sigma}\psi^*_{\sigma}(x)(i\partial_t-\xi_{\hat {p}})\psi_{\sigma}(x)+\sum_q\phi_q^*D^{-1}(\partial_t,q)\phi_q-\int{dx}\sum_{\sigma}\big\{g\psi^{*}_{\sigma}(x)\psi_{\sigma}(x)[\phi(x)+\phi^*(x)]+h.c.\big\}\Big\},
\end{equation}
where $\psi_{\sigma}(x)$ and $\phi(x)$ are the electron and phonon field operators, respectively; ${\hat p}=-i\hbar\partial_x$ denotes the momentum operator; $D(\partial_t,q)=\frac{\omega_q}{(i\partial_t)^2-\omega_q^2}$ represents the Green function of free phonons.

As first proposed in Ref.~\cite{lee1974conductivity} and developed in Refs.~\cite{hayashi1996topological,hayashi2000ginzburg,gruner1985charge,gruner1988dynamics}, one can single out the phonons around the wave number $\pm2k_F$ for the Peierls instability and obtain the action:
\begin{eqnarray}
  S&=&\int{dt}\Big\{\int{dx}\sum_{\sigma}\psi^*_{\sigma}(x)(i\partial_t-\xi_{\hat {p}})\psi_{\sigma}(x)+{\sum_q}'\phi_{Q+q}^*D^{-1}(\partial_t,Q+q)\phi_{Q+q}+{\sum_q}'\phi_{-Q+q}^*D^{-1}(\partial_t,-Q+q)\phi_{-Q+q}\nonumber\\
  &&\mbox{}-{\sum_{q}}'\sum_{\sigma}\int{dx}g[\psi^{*}_{\sigma}(x)\psi_{\sigma}(x)e^{iQx}(\phi_{Q+q}+\phi_{-Q-q}^{*})e^{iqx}+\psi^{*}_{\sigma}(x)\psi_{\sigma}(x)e^{-iQx}(\phi_{-Q+q}+\phi_{Q-q}^{*})e^{iqx}+h.c.]\Big\}\nonumber\\
  &\approx&\int{dt}\Big\{\int{dx}\sum_{\sigma}\psi^*_{\sigma}(x)(i\partial_t-\xi_{\hat {p}})\psi_{\sigma}(x)+{\sum_q}'\phi_{Q+q}^*D^{-1}(\partial_t,Q)\phi_{Q+q}+{\sum_q}'\phi_{-Q+q}^*D^{-1}(\partial_t,-Q)\phi_{-Q+q}\nonumber\\
  &&\mbox{}-{\sum_{q}}'\sum_{\sigma}\int{dx}g[\psi^{*}_{\sigma}(x)\psi_{\sigma}(x)e^{iQx}(\phi_{Q+q}+\phi_{-Q-q}^{*})e^{iqx}+\psi^{*}_{\sigma}(x)\psi_{\sigma}(x)e^{-iQx}(\phi_{-Q+q}+\phi_{Q-q}^{*})e^{iqx}+h.c.]\Big\}. \label{s3}
\end{eqnarray}
Here, the summation ${\sum_q}'$  of $q$ is restricted to long-wavelength regions, and we have neglected $q$-dependence of $\omega_{{\pm}Q+q}$. Following the mean-field treatment proposed in Refs.~\cite{lee1974conductivity,hayashi1996topological,hayashi2000ginzburg,gruner1985charge,gruner1988dynamics}, one can describe the lattice distortion by a complex order parameter:
\begin{equation}
\Delta(x)=g{\sum_q}'\langle{\phi_{Q+q}+\phi_{-Q-q}^*}\rangle{e^{iqx}},  
\end{equation}
with $\langle...\rangle$ being statistical average.  {{This approach explicitly takes the lattice distortion (phonon condensation) as the driving mechanism of CDW formation, and the corresponding lattice displacement is
\begin{eqnarray}
  u(x)&=&\sum_{q}\frac{1}{\sqrt{2M\omega_{q}}}e^{iqx}(b_{q}+b^{\dagger}_{-q})
  \Rightarrow\langle{u(x)}\rangle=\sum_{q}\frac{1}{\sqrt{2M\omega_{q}}}e^{iqx}(\langle{b_{q}}\rangle+\langle{b^{\dagger}_{-q}}\rangle)=\frac{1}{g\sqrt{2M\omega_{Q}}}|\Delta_0|\cos(Qx+\theta_0),
\end{eqnarray}
showing that the phonon condensate at $Q$ directly determines a periodic lattice modulation. }}

The electron field operator $\psi_{\sigma}(x)$ can be decomposed into the right- and left-moving components~\cite{hayashi1996topological,hayashi2000ginzburg}:
\begin{equation}
  \psi_{\sigma}(x)=\sum_{k}[e^{i(k_F+k)x}\mathcal{R}_{k\sigma}(R)+e^{i(-k_F+k)x}\mathcal{L}_{k\sigma}(R)],
\end{equation}
where $\mathcal{R}_{k\sigma}(R)$ and $\mathcal{L}_{k\sigma}(R)$ represent the right- and left-moving electron wave-functions, respectively, and $R$ denotes the center-of-mass coordinate associated with the long-wavelength fluctuations. Using this decomposition, the action of the microscopic model for conventional CDWs can be written as
\begin{equation}
  S=\int{dt}d{R}\sum_{k\sigma}\Psi_{k\sigma}^{\dagger}(R)\left(\begin{array}{cc}
    i\partial_t-\xi_{k_F+k-i\hbar\partial_R} & -\Delta(R) \\
    -\Delta^*(R) & i\partial_t-\xi_{-k_F+k-i\hbar\partial_R}
    \end{array}\right)\Psi_{k\sigma}(R)+\Delta^*(R)\Big[\frac{(i\partial_t)^2-\omega_Q^2}{2\omega_Qg^2}\Big]\Delta(R).\label{ss}
\end{equation}
Here, $\Psi_{k\sigma}^{\dagger}(R)=[\mathcal{R}^*_{k\sigma}(R),\mathcal{L}^*_{k\sigma}(R)]$. It should be emphasized that in the absence of the ${\Delta^*(i\partial_t^2)\Delta}/{(2\omega_Qg^2)}$ term, this action is formally analogous to that of a BCS superconductor~\cite{abrikosov2012methods,bardeen1957theory}. The inclusion of this term in CDWs modifies the dynamic properties of the order parameter, i.e., energy spectra of the amplitude and phase modes, as analytically demonstrated by Lee, Rice and Anderson in Ref.~\cite{lee1974conductivity}, leading to distinct differences from the superconducting case. {{Clearly, through the electron–phonon interaction, the lattice distortion
opens a gap in the electronic spectrum and produces a periodic
modulation of the electronic charge density here. In this CDW model, the lattice distortion is the cause, while the charge modulation and electronic gap opening are the consequences.}}

\section{Derivation of the effective action of CDWs}

In this section, we derive the effective action of CDWs by treating the CDW gap and phase on an equal footing within the standard path-integral framework. For the convenience, we set $\hbar=1$ here and hereafter. Specifically, a general complex order parameter of CDWs reads $\Delta(R)=(|\Delta_0|+\delta|\Delta|)\exp[{i\theta_0+i\delta\theta(R)}]$, where $|\Delta|=|\Delta_0|+\delta|\Delta|$ and $\theta=\theta_0+\delta\theta(R)$. Here, $|\Delta_0|$ and $\theta_0$ correspond to the CDW gap and phase, respectively, while $\delta|\Delta|$ and $\delta\theta(R)$ represent the amplitude and phase fluctuations. 

Applying the chiral transformations~\cite{hayashi2000ginzburg}:
\begin{equation}
\mathcal{R}_{k\sigma}\rightarrow\mathcal{R}_{k\sigma}\exp[{i\theta_0/2+i\delta\theta(R)}/2],~~~~~\mathcal{L}_{k\sigma}\rightarrow\mathcal{L}_{k\sigma}\exp[-{i\theta_0/2-i\delta\theta(R)}/2],
\end{equation}
and neglecting the trivial terms, the action becomes
\begin{eqnarray}\label{sss}
  S&=&\int{dt}dR\Big\{\sum_{k\sigma}\Psi_{k\sigma}^{\dagger}(R)\Big[i\partial_t-\frac{v_F\partial_R\delta\theta}{2}-\frac{(\partial_R\delta\theta)^2}{8m}-\varepsilon_k\tau_3-\frac{\partial_t\delta\theta}{2}\tau_3-|\Delta_0|\tau_1-\delta|\Delta|\tau_1\Big]\Psi_{k\sigma}(R)-\frac{\omega_Q(|\Delta_0|+\delta|\Delta|)^2}{2g^2}\nonumber\\
  &&\mbox{}+\frac{|\Delta_0|^2(\partial_t\delta\theta)^2}{2\omega_Qg^2}+\frac{\delta|\Delta|(i\partial_t^2)\delta|\Delta|}{2g^2\omega_Q}-\frac{\delta|\Delta|2i\partial_t\delta\theta\partial_t\delta|\Delta|}{2g^2\omega_Q}\Big\}\nonumber\\
  &=&\int{dt}dR\Big\{\sum_{k\sigma}\Psi_{k\sigma}^{\dagger}(R)(G_0^{-1}\!-\!\Sigma)\Psi_{k\sigma}(R)\!-\!\frac{\omega_Q(|\Delta_0|\!+\!\delta|\Delta|)^2}{2g^2}\!+\!\frac{|\Delta_0|^2(\partial_t\delta\theta)^2}{2\omega_Qg^2}\!+\!\frac{\delta|\Delta|(i\partial_t^2)\delta|\Delta|}{2g^2\omega_Q}\!-\!\frac{\delta|\Delta|2i\partial_t\delta\theta\partial_t\delta|\Delta|}{2g^2\omega_Q}\Big\},~~~ 
\end{eqnarray}
where $\tau_i$ denotes the Pauli matrices in right- and left-moving space; the Green function is determined by $G^{-1}_0=i\partial_t-{v_Fp_s}-\varepsilon_k\tau_3-|\Delta_0|\tau_1$, with $p_s=\partial_R\delta\theta/{2}$, and the self-energy reads $\Sigma={p_s^2}/({2m})+{\partial_t\delta\theta}\tau_3/2-\delta|\Delta|\tau_1$. It should be emphasized that unlike in superconductors, where the amplitude and phase modes are decoupled due to the particle-hole symmetry~\cite{pekker2015amplitude,yang2019gauge,yang2021theory,sun2020collective}, the amplitudon and phason in CDWs are intrinsically coupled due to the presence of the last term in Eq.~(\ref{sss}), which arises from the Green function of free phonons, i.e., last term in Eq.~(\ref{ss}).

Then, through the standard integration over the Fermi field in the Matsubara representation~\cite{schrieffer1964theory,yang2021theory,yang2024diamagnetic,yang2023optical,sun2020collective}, one has
\begin{equation}
  S=\int{dt}dR\Big\{{\rm {\bar T}r}\ln{G^{-1}_0}-\sum_n^{\infty}\frac{1}{n}{\rm {\bar T}r}[(\Sigma{G_0})^n]-\frac{\omega_Q(|\Delta_0|+\delta|\Delta|)^2}{2g^2}+\frac{|\Delta_0|^2(\partial_t\delta\theta)^2}{2\omega_Qg^2}+\frac{\delta|\Delta|(i\partial_t^2)\delta|\Delta|}{2g^2\omega_Q}-\frac{\delta|\Delta|2i\partial_t\delta\theta\partial_t\delta|\Delta|}{2g^2\omega_Q}\Big\},
\end{equation}
where the Green function in the Matsubara representation: 
\begin{equation}\label{GreenfunctionMr}
G_0(ip_n,k)=\frac{ip_n\tau_0-{p}_s{v_F}\tau_0+\varepsilon_k\tau_3+|\Delta_0|\tau_1}{(ip_n-E_{k}^+)(ip_n-E_{k}^-)},  
\end{equation}  
with $E_{k}^{\pm}=v_Fp_s\pm\sqrt{\varepsilon_k^2+|\Delta_0|^2}=v_Fp_s\pm{E_k}$. Further keeping the lowest two orders (i.e., $n=1$ and $n=2$) and neglecting the trivial terms lead to the final result: $S_{\rm eff}=S_{\rm eff}^e+S_{\rm eff}^{A}$, with 
the effective action of the CDW gap and phase:
\begin{equation}\label{ace}
  S_{\rm eff}^e=\int{dt}dR\Big\{\sum_{ip_n,k,\sigma}\ln[(ip_n-E_k^+)(ip_n-E_k^-)]-\frac{p_s^2}{2m}\chi_0-\big(\frac{\partial_t\delta\theta}{2}\big)^2\chi_{33}+\frac{|\Delta_0|^2}{2\omega_Qg^2}(\partial_t\delta\theta)^2-\frac{\omega_Q}{2g^2}|\Delta_0|^2\Big\},
\end{equation}
and the effective action of amplitudon:
\begin{equation}\label{aca}
S^A_{\rm eff}=-\int{dt}dR\Big\{\big(\chi_1+\frac{\omega_Q}{2g^2}2|\Delta_0|\big)\delta|\Delta|+\chi_{11}(\delta|\Delta|)^2+\frac{\omega_Q}{2g^2}(\delta|\Delta|)^2+\frac{\delta|\Delta|\partial_t^2\delta|\Delta|}{2g^2\omega_Q}+\frac{\delta|\Delta|2i\partial_t\delta\theta\partial_t\delta|\Delta|}{2g^2\omega_Q}\Big\}.  
\end{equation}
Here, the correlation coefficients are given by (similar to the ones in BCS superconductors~\cite{yang2023optical,yang2021theory})
\begin{eqnarray}
  \chi_0&=&\sum_{ip_n,k,\sigma}{\rm Tr}[G_0(ip_n,k)\tau_0]=n=\sum_{\sigma}\frac{2k_F}{2\pi}, \label{chi0}\\
  \chi_1&=&\sum_{ip_n,k,\sigma}{\rm Tr}[G_0(ip_n,k)\tau_1]=2\sum_{ip_n,k,\sigma}\frac{|\Delta_0|}{(ip_n-E_{k}^+)(ip_n-E_{k}^-)}=2\sum_{k\sigma}|\Delta_0|\frac{f(E_k^+)-f(E_k^-)}{2E_k},\\
  \chi_{33}&=&\frac{1}{2}\sum_{ip_n,k,\sigma}{\rm Tr}[G_0(ip_n,k)\tau_3G_0(ip_n,k)\tau_3]=\sum_{ip_n,k,\sigma}\frac{(ip_n-{p}_s{v_F})^2+\varepsilon_k^2-|\Delta_0|^2}{[(ip_n-{p}_s{v_F})^2-\varepsilon_k^2-|\Delta_0|^2]^2}\nonumber\\
  &=&\sum_{ip_n,k,\sigma}\frac{\Big[1+\frac{2\varepsilon_k^2}{(ip_n-{p}_s{v_F})^2-\varepsilon_k^2-|\Delta_0|^2}\Big]}{(ip_n-{p}_s{v_F})^2-\varepsilon_k^2-|\Delta_0|^2}=\frac{1}{2}\sum_{ip_n,k,\sigma}\partial_{\varepsilon_k}\{{\rm Tr}[G_0(ip_n,k)\tau_3]\}=-\frac{1}{\pi{v_F}}=-D_0.
\end{eqnarray}  
The correlation coefficient $\chi_{11}$, related to the dynamic properties of the amplitudon, will be discussed later.

\section{CDW pinning  by impurities}

In this section, we introduce the CDW pinning term, which was proposed and developed by Fukuyama and Lee~\cite{fukuyama1978dynamics} as well as by Rice, Whitehouse and Littlewood~\cite{rice1981impurity}. The interaction of CDW with impurity potential $V(x-R_i)$ located at $R_i$ is written as~\cite{fukuyama1978dynamics,rice1981impurity}
\begin{equation}
H_i=\sum_i\int{dx}\rho(x)V(x-R_i).  
\end{equation}
Assuming a short-range impurity potential $V(x-R_i)=V_0\delta(x-R_i)$, the action related to the impurity pining term on the phase of CDW is written as
\begin{equation}
  S_i=\int{dt}\int{dR}\Big\{-V_0nf_s\sum_i\cos[QR_i+\theta_0(R_i)]+\frac{V_0nf_s}{2}\sum_i[\theta(R_i)-\theta_0(R_i)]^2\cos[QR_i+\theta_0(R_i)]\Big\},  \label{SSI}
\end{equation}
in which the first term describes the part of the equilibrium phase and the second term describes the dynamics of the phase. Here, $f_s$ denotes the condensation ratio~\cite{hayashi1996topological,hayashi2000ginzburg}. 

In the case of strong pinning, where the impurity potential is either strong or the impurity concentration is dilute, the CDW distorts to maximize the interaction with the impurity potential~\cite{fukuyama1978dynamics,rice1981impurity}, resulting in the relation $\langle\cos[QR_i+\theta_0(R_i)]\rangle\approx-1$. In contrast, for the weak pinning case, where the impurity potential is weak or the impurity concentration is high, as described in Ref.~\cite{rice1981impurity}, the phase $\theta_0(R)$ varies slowly, and the total phase $QR_i+\theta_0(R_i)$ becomes nearly random. This leads to the approximation $\langle\cos[QR_i+\theta_0(R_i)]\rangle\approx-\frac{L(n_i L_0)^{1/2}}{L_0}$, where $L$ is the length of the CDW chain and $L_0$ is the domain length. By minimizing the impurity potential energy and elastic energy, Ref.~\cite{rice1981impurity} reports the expression for $L_0$ as $L_0^{-1}=[(\frac{3V_0\rho_0}{{\pi}v_F})^2n_i]^{1/3}$. 

Consequently, the action associated with the impurities can be written as 
\begin{equation}\label{s}
  S_i=\int{dt}\int{dR}\Big\{-f_sr_i^2\Big(\frac{\delta\theta}{2}\Big)^2+2f_sr_i^2\Big\},  
\end{equation}
where
\begin{equation}
  r_i^2 = 
  \begin{cases}
    2nn_iV_0 & \text{strong pinning}, \\
    2nn_iV_0\sqrt{n_iL_0}\frac{L}{L_0} & \text{weak pinning},
  \end{cases}
\end{equation}
with $n_i$ being the impurity density.  While the last term in Eq.~(\ref{s}) is trivial for phase dynamics, the first term in both the strong and weak pinning cases introduces an excitation gap (i.e., an effective mass) in the phase mode (phason). As a result, in the low-temperature limit, impurities can always pin the CDW. Thus, the low-dimensional CDWs are not constrained by the Mermin–Wagner theorem, which forbids the formation of long-range order in low-dimensional systems.

Notably, it is established in the literature that the enhanced thermal phase fluctuations at elevated temperatures can weaken the pinning. Within the self-consistent field approximation, as assumed/proposed in several early works~\cite{maki1986thermal,rice1981impurity}, this effect can be taken into account by considering the phason softening as temperature increases, i.e., by replacing $r_i^2$ in Eq.~(\ref{s}) with a self-consistent temperature-dependent one $r_i^2(T)=r_i^2(T=0)\exp[-\xi^2(\nabla_R\delta\theta/2)^2]=r_i^2(T=0)\exp(-\xi^2p_s^2)$ with $\xi(T)\propto{\hbar{v_F}/|\Delta_0(T)|}$ being the coherence length of CDWs.

\section{Derivation of the self-consistent phase-transition theory}

In this section, we derive the self-consistent phase-transition theory of 1D CDWs. Specifically, with Eq.~(\ref{ace}), through the the variation with respect to the CDW gap at equilibrium, i.e., $\partial_{|\Delta_0|}S_{\rm eff}^e=0$, one has the gap equation:
\begin{eqnarray}\label{GE}
-\!\!\!\sum_{ip_n,k,\sigma}\frac{2|\Delta_0|}{(ip_n\!-\!E_k^+)(ip_n\!-\!E_k^-)}\!-\!2|\Delta_0|\frac{\omega_Q}{2g^2}\!=\!0~~\Rightarrow~~\frac{\omega_Q}{2g^2}\!=\!-\!\!\!\sum_{ip_n,k,\sigma}\frac{1}{(ip_n\!-\!E_k^+)(ip_n\!-\!E_k^-)}\!=\!-\frac{1}{{\pi}v_F}\int\frac{f(E_k^+)\!-\!f(E_k^-)}{2E_k}{d\varepsilon_k}.~~
\end{eqnarray}  

With Eqs.~(\ref{ace}) and~(\ref{s}), through the Euler-Lagrange equation of motion with respect to the phase fluctuations in the presence of impurities, i.e., $\partial_{\mu}[\frac{\partial{(S^e_{\rm eff}+S_i)}}{\partial(\partial_{\mu}\delta\theta/2)}]=\partial_{\delta\theta/2}(S^e_{\rm eff}+S_i)$, the equation of motion of the phase fluctuations reads
\begin{eqnarray}
  &&  -~\partial_R\Big[\sum_{ip_n,k,\sigma}\frac{v_F(ip_n-v_Fp_s)}{(ip_n-E_k^+)(ip_n-E_k^-)}\Big]+\Big[-\frac{\partial_R^2}{m}\chi_0+\frac{1}{\pi{v_F}}\Big(1+\frac{4|\Delta_0|^2}{\lambda\omega^2_Q}\Big)\partial_t^2+f_sr_i^2\Big]\frac{\delta\theta}{2}=0\nonumber\\
  &&\Rightarrow~~\Big\{\frac{1}{\pi{v_F}}\Big(1+\frac{4|\Delta_0|^2}{\lambda\omega^2_Q}\Big)\partial_t^2-\Big[\frac{\chi_0}{m}+2v_F^2\sum_{k}\partial_{E_k}\Big(\frac{f(E_k^+)-f(E_k^-)}{2}\Big)\Big]\partial_R^2+f_sr_i^2\Big\}\frac{\delta\theta}{2}=0,  
\end{eqnarray}
which leading to the result:
\begin{equation}
D^*_0\Big[\partial_t^2-\frac{m}{m^*}v_F^2f_s\partial_R^2+\frac{m}{m^*}f_sm_{\rm P}^2(T)\Big]\frac{\delta\theta}{2}=0.  
\end{equation}
Here, the renormalized density of states $D_0^*=D_0{m^*}/{m}$, and $\frac{m^*}{m}=\Big(1+\frac{4|\Delta_0|^2}{\lambda\omega^2_Q}\Big)$, exactly the same as the one derived in previous works~\cite{gruner1985charge,gruner1988dynamics,lee1974conductivity}; the phason mass (excitation gap) $m_{\rm P}$ due to impurities is determined by $m^2_{\rm P}(T)=r_i^2(T)\pi{v_F}$; the condensation ratio with the help of Eq.~(\ref{chi0}) is derived as
\begin{eqnarray}
  f_s&=&{\pi}v_F\Big[\frac{\chi_0}{mv_F^2}+2\sum_{k}\partial_{E_k}\Big(\frac{f(E_k^+)-f(E_k^-)}{2}\Big)\Big]=1+\int{d\varepsilon_k}\partial_{E_k}\Big(\frac{f(E_k^+)-f(E_k^-)}{2}\Big)\nonumber\\
  &=&-\int{d\varepsilon_k}\partial_{\varepsilon_k}\Big(\frac{f(E_k^+)-f(E_k^-)}{2E_k}\varepsilon_k\Big)+\int{d\varepsilon_k}\partial_{E_k}\Big(\frac{f(E_k^+)-f(E_k^-)}{2}\Big)\nonumber\\
  &=&\int{d\varepsilon_k}\Big[\Big(1-\frac{\varepsilon_k^2}{E_k^2}\Big)\partial_{E_k}\Big(\frac{f(E_k^+)-f(E_k^-)}{2}\Big)-\frac{|\Delta_0|^2}{E_k^3}\frac{f(E_k^+)-f(E_k^-)}{2}\Big]=\int{d\varepsilon_k}\frac{|\Delta_0|^2}{E_k}\partial_{E_k}\Big(\frac{f(E_k^+)-f(E_k^-)}{2E_k}\Big),\label{fs}
\end{eqnarray}
analogous to the superfluid ratio in BCS superconductors~\cite{yang2018gauge,yang2019gauge,yang2020influence,yang2021theory,yang2023optical,yang2024optical}.  As a result, one finds $f_s\approx1+\int{d\varepsilon_k}\partial_{E_k}f(E_k)$ at low-temperature limit and $f_s=1$ at $T=0$, and $f_s=\frac{7|\Delta_0|^2\zeta(3)}{16(\pi{T})^2}$ for $|\Delta_0|\rightarrow0$, where $\zeta(n)$ denotes the Riemann zeta function~\cite{yang2018gauge}.

Consequently, in momentum space, the equation of motion of phason becomes
\begin{equation}\label{EOM}
D^*_0\Big[\partial_t^2+\frac{m}{m^*}f_sv_F^2q^2+\frac{m}{m^*}f_sm_{\rm P}^2(T)\Big]\frac{\delta\theta_q}{2}=0,  
\end{equation}
from which one obtains the phason energy spectrum:
\begin{equation}\label{phasonenergy}
  \Omega_{\rm P}^2(q)=\frac{m}{m^*}f_sm^2_{\rm P}(T)+\frac{m}{m^*}{f_sq^2v_F^2},
\end{equation}
which, in the absence of CDW pinning, recovers the one derived in the previous work~\cite{lee1974conductivity}.\\

From the equation of motion of the phase fluctuations, one can derive the thermal phase fluctuations through the fluctuation dissipation theorem or within Matsubara formalism. The two methods are equivalent and lead to exactly the same result.   

{\sl Fluctuation dissipation theorem.---}Considering the thermal fluctuations, the dynamics of the phase from Eq.~(\ref{EOM}) is given by
\begin{equation}
\big[D^*_0\Omega_{\rm P}^2(q)-D^*_0\omega^2+iD^*_0\omega{\gamma}\big]\frac{\delta\theta({\omega,{q}})}{2}={J}_{\rm th}(\omega,{q}).    
\end{equation}  
Here, we have introduced a thermal field ${J}_{\rm th}(\omega,{q})$ that obeys the fluctuation-dissipation theorem~\cite{Landaubook}:
\begin{equation}
\langle{J}_{\rm th}(\omega,{q}){J}^*_{\rm th}(\omega',{q}')\rangle=\frac{(2\pi)^2D^*_0\gamma\omega\delta(\omega-\omega')\delta({q-q'})}{\tanh(\beta\omega/2)},   
\end{equation}
and $\gamma=0^+$ is a phenomenological damping constant. From this dynamics, the average of the phase fluctuations is given by
\begin{eqnarray}
  \langle{p_s^2}\rangle&=&\int\frac{d\omega{d\omega'}d{q}d{q'}}{(2\pi)^4}\frac{({q}{q'})\langle{J}_{\rm th}(\omega,{q}){J}^*_{\rm th}(\omega',{q}')\rangle}{(D^*_0)^2\big[\big(\omega^2\!-\!\Omega_{\rm P}^2(q)\big)\!-\!i\omega\gamma\big]\big[({\omega'}^2\!-\!\Omega_{\rm P}^2(q')\big)\!+\!i\omega'\gamma\big]}=\int\frac{d\omega{d{q}}}{D^*_0(2\pi)^2}\frac{q^2\gamma\omega/\tanh(\beta\omega/2)}{\big[\omega^2\!-\!\Omega_{\rm P}^2(q)\big]^2+\omega^2\gamma^2}\nonumber\\
&=&\int\frac{{d{q}}}{2\pi}\frac{q^2}{2D^*_0\Omega_{\rm P}(q)}[2n_B\big(\Omega_{\rm P}(q)\big)+1].\label{EEEE}
\end{eqnarray}

{\sl Matsubara formalism.---}Within the Matsubara formalism, by mapping into the imaginary-time space, from Eq.~(\ref{EOM}),  the thermal phase fluctuation reads~\cite{abrikosov2012methods}
\begin{eqnarray}
  \langle{p_s^2}\rangle&\!=\!&\int\frac{d{q}}{(2\pi)}q^2\Big[\Big\langle\Big|\frac{\delta\theta^*(\tau,{q})}{2}\frac{\delta\theta(\tau,{q})}{2}e^{-\int^{\beta}_{0}d\tau{d{q}}D^*_0{\delta\theta^*(\tau,{q})}(\Omega_{\rm P}^2-\partial_{\tau}^2){\delta\theta(\tau,{q})}/4}\Big|\Big\rangle\Big]\nonumber\\
  &\!=\!&\int\frac{d{q}}{(2\pi)}q^2\Big[\frac{1}{\mathcal{Z}_0}{\int}D\delta\theta{D\delta\theta^*}\frac{\delta\theta^*(\tau,{q})}{2}\frac{\delta\theta(\tau,{q})}{2}e^{-\int^{\beta}_{0}d\tau{d{q}}D^*_0{\delta\theta^*(\tau,{q})}(\Omega_{\rm P}^2-\partial_{\tau}^2){\delta\theta(\tau,{q})}/4}\Big]\nonumber\\
   &=&\!\!\!\int\frac{d{q}}{(2\pi)}q^2\frac{1}{\mathcal{Z}_0}{\int}D\delta\theta{D\delta\theta^*}\delta_{J^*_{q}}\delta_{J_{q}}e^{-\int^{\beta}_{0}d\tau{d{q}}[D^*_0{\delta\theta^*(\tau,{q})}(\Omega_{\rm P}^2-\partial_{\tau}^2){\delta\theta(\tau,{q})}/4+J_{q}\delta\theta({q})/2+J^*_{q}\delta\theta^*({q})/2]}\Big|_{J=J^*=0}\nonumber\\
  &=&\!\!\!\int\frac{d{q}}{(2\pi)}\frac{q^2}{D^*_0}\delta_{J^*_{q}}\delta_{J_{q}}\exp\Big\{-\int_0^{\beta}{d\tau}\sum_{q'}J_{q'}\frac{1}{\partial_{\tau}^2-\Omega_{\rm P}^2}J^*_{q'}\Big\}\Big|_{J=J^*=0}=-\int\frac{d{q}}{2\pi}\frac{1}{\beta}\sum_{\omega_n}\frac{q^2}{D^*_0}\frac{1}{(i\Omega_n)^2\!-\!\Omega_{\rm P}^2}\nonumber\\
 &=&\int\frac{{d{q}}}{2\pi}\frac{q^2}{2D^*_0\Omega_{\rm P}(q)}[2n_B(\Omega_{\rm P})+1],\label{mf}
\end{eqnarray}
which is exactly the same as the one in Eq.~(\ref{EEEE}) obtained via the fluctuation dissipation theorem. Here, $\Omega_n=2n\pi{T}$ represents the Bosonic Matsubara frequencies; ${J_{q}}$ denotes the generating functional and $\delta{J_{q}}$ stands for the functional derivative.\\

For the accuracy of the quantitative description, all the model parameters used in our analysis are taken from the experimentally measured ones in the low-temperature limit.  The experimentally measured observables in the low-temperature limit in fact  reflect a renormalized ground state that already includes zero-point fluctuations. {In general, one cannot simultaneously extract low temperature parameters from experiments and include
explicit vacuum fluctuations without double counting.}  Following the treatment of the zero-point oscillations in the quantum field theory by integrating it to the ground state~\cite{peskin2018introduction},  here we directly subtract the zero-point oscillations in Eq.~(\ref{EEEE}) or Eq.~(\ref{mf}) and obtain the sole thermal phase fluctuations:
\begin{eqnarray}\label{TP}
  \langle{p_s^2}\rangle &=&\int\frac{{d{q}}}{2\pi}\frac{q^2n_B(\Omega_{\rm P})}{D^*_0\Omega_{\rm P}(q)}.
\end{eqnarray}
{This treatment should be viewed as a controlled heuristic approximation that extracts effective low-temperature parameters directly from experiments to avoid double counting of explicit vacuum fluctuations.} Then, the remaining temperature dependence is governed solely by the thermal fluctuations.  Consequently, we arrive at the developed self-consistent phase-transition theory, consisting of the CDW gap equation in Eq.~(\ref{fs}), thermal phase fluctuations in Eq.~(\ref{TP}) and CDW-condensation ratio in Eq.~(\ref{GE}), as summarized in the main text.

\section{Derivation of the energy spectrum of amplitudon}

In this section, we derive the energy spectrum of the amplitudon by using $S_{\rm eff}^A$ from Eq.~(\ref{aca}). The first term in Eq.~(\ref{aca}) vanishes due to the CDW gap equation. Then,  through the Euler-Lagrange equation of motion with respect to the amplitude fluctuations, i.e., $\partial_{t}[\frac{\partial{S^A_{\rm eff}}}{\partial(\partial_{t}\delta|\Delta|)}]=\partial_{\delta|\Delta|}S^A_{\rm eff}$, the equation of motion of the amplitudon reads
\begin{equation}
 \big(\chi_{11}+\frac{\omega_Q}{2g^2}\big)\delta|\Delta|+\frac{\partial^2_t\delta|\Delta|}{2g^2\omega_Q}=0,  
\end{equation}
where the dynamic coefficient is derived as (similar to the one in BCS superconductors~\cite{yang2023optical,yang2024diamagnetic})
\begin{eqnarray}
  &&\chi_{11}(i\Omega_n)\!+\!\frac{\omega_Q}{2g^2}=\frac{1}{2}\sum_{ip_n,k,\sigma}{\rm Tr}[G_0(ip_n+i\Omega_n,k)\tau_1G_0(ip_n,k)\tau_1]\!+\!\frac{\omega_Q}{2g^2}\nonumber\\
  &&\mbox{}=\sum_{ip_n,k,\sigma}\frac{(ip_n\!+\!v_Fp_s\!+\!i\Omega_n)(ip_n\!+\!v_Fp_s)+|\Delta_0|^2-\varepsilon^2_{k}}{[(ip_n\!+\!v_Fp_s\!+\!i\Omega_n)^2\!-\!E^2_{k}][(ip_n\!+\!v_Fp_s)^2\!-\!E^2_{k}]}\!+\!\frac{\omega_Q}{2g^2}=\!\!\sum_{ip_n,k,\sigma}\Big\{\frac{4|\Delta_0|^2-(ip_n\!+\!v_Fp_s\!+\!i\Omega_n-ip_n\!-\!v_Fp_s)^2}{2[(ip_n\!+\!v_Fp_s\!+\!i\Omega_n)^2-E^2_{k}][(ip_n\!+\!v_Fp_s)^2\!-\!E^2_{k}]}\!\nonumber\\
  &&\mbox{}+\!\frac{1}{2[(ip_n\!+\!v_Fp_s)^2\!-\!E^2_{k}]}\!+\!\frac{1}{2[(ip_n\!+\!v_Fp_s\!+\!i\Omega_n)^2\!-\!E^2_{k}]}-\!\frac{1}{[(ip_n\!+\!v_Fp_s)^2\!-\!E^2_{k}]}\!\approx\!\!\!\!\sum_{ip_n,k,\sigma}\frac{4|\Delta_0|^2\!-\!(i\Omega_n)^2}{2[(ip_n\!+\!v_Fp_s)^2\!-\!E^2_{k}]^2}\nonumber\\
  &&\mbox{}=({4|\Delta_0|^2\!-\!(i\Omega_n)^2})\sum_{k,\sigma}\Big[\frac{\partial_{E_k}\big(f(E_k^+)\!-\!f(E_k^-)\big)}{8E_k^2}\!-\!\frac{f(E_k^+)\!-\!f(E_k^-)}{8E_{k}^3}\Big]\nonumber\\
  &&\mbox{}=\Big(1\!-\!\frac{(i\Omega_n)^2}{4|\Delta_0|^2}\Big)D_0\int{d\varepsilon_k}\frac{|\Delta_0|^2}{E_k}\partial_{E_k}\Big(\frac{f(E_k^+)-f(E_k^-)}{2E_{k}}\Big)=\Big(1\!-\!\frac{(i\Omega_n)^2}{4|\Delta_0|^2}\Big)D_0f_s,~~~~~ \label{c11}
\end{eqnarray}
where $\Omega_n=2n\pi{T}$ represents the Bosonic Matsubara frequencies.

As a result, in the real frequency space, the equation of motion of the amplitudon becomes
\begin{equation}
 \Big[\Big(1\!-\!\frac{\Omega^2}{4|\Delta_0|^2}\Big)D_0f_s-\frac{1}{2g^2\omega_Q}\Omega^2\Big]\delta|\Delta|=0~~~\Rightarrow~~~\Big[\frac{f_s\lambda\omega^2_Q}{1+{f_s\lambda\omega^2_Q}/({4|\Delta_0|^2})}\!-\!\Omega^2\Big]\delta|\Delta|=0,  
\end{equation}
leading to the energy spectrum of the amplitudon at the long-wavelength limit:
\begin{equation}
\Omega_A=\sqrt{\frac{f_s\lambda\omega^2_Q}{1+{f_s\lambda\omega^2_Q}/({4|\Delta_0|^2})}}.  
\end{equation}
At zero temperature with $f_s=1$ and $2|\Delta_0|\gg{\sqrt{\lambda}\omega_Q}$, $\Omega_{\rm A}$ reduces to $\sqrt{\lambda}\omega_Q$, the same as the one obtained in previous works at $T=0$~\cite{gruner1985charge,gruner1988dynamics,lee1974conductivity}. Conversely, if one artificially assumes a small $|\Delta_0|$, the derived $\Omega_{\rm A}$ approximates $2|\Delta_0|$, consistent with the predictions from the phenomenological Landau phase-transition theory~\cite{littlewood1981gauge,pekker2015amplitude,yang2024optical,yang2023optical} and analogous to the Higgs/amplitude mode (with excitation energy $\omega_H=2|\Delta_0|$) in superconductors~~\cite{littlewood1981gauge,pekker2015amplitude,matsunaga2013higgs,matsunaga2014light,yang2019gauge,shimano2020higgs,sun2020collective,yang2020theory}.

The damping term of the amplitudon that arises from the intrinsic coupling with the phason [i.e., the last term in Eq.~(\ref{aca})] requires a detailed microscopic scattering treatment (e.g., high-order diagrammatic technique~\cite{yang2024diamagnetic} or nonequilibrium Green-function approach~\cite{yang2020influence,yang2018gauge}), which is beyond the scope of the present study. This process should be described by the
three-boson (two amplitudon and one phason) scattering, and according to the
microscopic scattering mechanism for calculating the scattering probabilities of the boson-emission and -absorption process of the scattered particles~\cite{yang2015hole,yang2016spin,yang2025THz},  the damping rate of the amplitudon can be approximated as $\gamma(T)\propto2N_{\rm phason}(T)+1$ with $N_{\rm phason}(T)$ being the averaged phason number. As a result, in the specific simulations, the damping rate can be approximated as $\gamma(T)=\gamma_{th}(T)+\gamma_0$, where $\gamma_0$ is a temperature-independent constant, and $\gamma_{th}(T)=c_0\int\frac{dq}{2\pi}\frac{n_B(\Omega_{{\rm P}}(q))}{D^*_0\Omega_{{\rm P}}(q)}$ accounts for contributions from the thermal excitation of the phason, i.e., phason number. The equation of motion of the amplitudon then reads
\begin{equation}
(\Omega^2-i\Omega\gamma-\Omega^2_{\rm A})\delta|\Delta|=S_e,
\end{equation}
with $S_e$ being the source term by the external stimulus. 

 {While the amplitudon is not dipole-active in an ideal, translationally invariant crystal, in real pinned CDWs (studied here), impurities, defects, lattice imperfections, grain boundaries and surface effects inevitably breaks translational symmetry. Such symmetry breaking can lead to a finite polarization associated with amplitude fluctuations, allowing the amplitudon to acquire optical activity and produce a measurable resonance at the amplitude-mode frequency~\cite{gruner1988dynamics,DONOVAN1990721}.  Specifically, in standard CDW theory the electronic charge density takes the form
\begin{equation}
\rho(R)=\rho_0+\rho_1\cos[QR+\theta_0(R)], 
\qquad \rho_1\propto |\Delta| .
\end{equation}
Thus, an amplitude fluctuation
\begin{equation}
|\Delta(t)| = |\Delta_0| + \delta|\Delta(t)|
\end{equation}
directly induces a modulation of the charge density
\begin{equation}
\delta\rho(R,t)\propto \delta|\Delta(t)|\cos[QR+\theta_0(R)],
\end{equation}
and then, the macroscopic polarization takes the form:
\begin{equation}
P(t) = \int dR\; R\cdot{\delta\rho(R,t)}\propto\delta|\Delta(t)|\int dR\; R\cdot{\cos[QR+\theta_0(R)]}.
\end{equation}
In an ideal, perfectly translationally invariant crystal [$\theta_0(R)\equiv\text{const.}$], the spatial average of the modulation $\delta\rho(R,t)$ vanishes, yielding no net dipole moment. In real quasi-1D CDW material, spatial inhomogeneities (impurities, defects, lattice imperfections, grain boundaries, and domain structure) break translational symmetry, yielding a finite   spatial average $\langle\cos[QR+\theta_0(R)]\rangle$ [as shown by the derivation of Eq.~(\ref{s}) from Eq.~(\ref{SSI})].  Then, the charge-density modulation no longer averages to zero and generates a non-vanishing time-dependent macroscopic polarization, acting as a radiating dipole and producing coherent far–field THz emission. Thus, once translational symmetry is broken (as is always the case in a pinned CDW),  the amplitude oscillation acquires an induced dipole moment and becomes optically active, making coherent THz emission possible.}

\begin{center}
\begin{table}[htb]
  \caption{Specific parameters used in our developed thermodynamic theory for quasi-1D CDW material (TaSe$_4$)$_2$I. For (TaSe$_4$)$_2$I, the tunneling spectroscopic measurements~\cite{huang2021absence,gooth2019axionic,forro1987hall,wang1983charge} reported a CDW gap of $\sim200~$meV at low temperatures, and the frequency $\omega_{Q}\approx$0.25-0.26~THz as measured using inelastic neutron scattering~\cite{Lorenzo_1998,fujishita1986neutron}.
    The cutoff energy $\omega_D$ is determined by the upper band bound ($0.2~$eV above the Fermi level) from first-principles calculations~\cite{lin2024unconventional}, as the Dirac point is located $\sim0.4~$eV below the Fermi level as reported by several previous studies~\cite{lin2024unconventional,huang2021absence,yi2021surface,tournier2013electronic}. The coupling constant $\lambda$ is then determined by fitting the experimental zero-temperature CDW gap $|\Delta_0(T=0)|$ using CDW gap equation. For thermal phase fluctuations, the cutoff $q_c$ is determined by $\pi/(2q_c)=d_{110}\approx0.69~$nm~\cite{huang2021absence}, interplane distance along the [110] direction, thereby restricting the thermal phase fluctuations to wavelengths no shorter than the interchain separation. This choice ensures that only physically relevant long-wavelength phason modes are retained. The damping parameters of the amplitudon, $\gamma_0$ and $c_0$, are extracted from experimental data in Ref.~\cite{nguyen2023ultrafast} via fitting the broadening of the THz-emission coherent signal at $7$~K.  }  
\renewcommand\arraystretch{1.5}   \label{TASEI}
\begin{tabular}{l l l l l l l l}
    \hline
    \hline
    model parameters of CDW gap equation&\quad\quad~$|\Delta_0(T=0)|~$(meV)&\quad\quad~$\omega_{2k_F}$(meV)&\quad\quad\quad~$\omega_D~$(meV)\\   
    &\quad\quad\quad\quad200&\quad\quad\quad$1.0678$&\quad\quad\quad\quad~$200$\\
     \hline
    model parameters of phase fluctuations&\quad\quad~$v_F~$(cm/s)~\cite{brutting1995dc}&\quad\quad\quad~$m_{\rm P}(T=0)~$\\   &\quad\quad\quad$6.59\times10^7$&\quad\quad\quad$0.9\hbar{q_cv_F}$\\
     \hline
    damping parameters of amplitudon&\quad\quad$1/\gamma_0$~(ps)&\quad\quad$1/c_0$~(ps)\\   
    &\quad\quad11.468&\quad\quad1.023\\
    \hline
    \hline
\end{tabular}\\
\end{table}
\end{center}

\section{Simulation parameters}

In our numerical simulations to solve the developed self-consistent phase-transition theory  1D CDWs,  we self-consistently calculate the gap equation [Eq.~(\ref{fs})], thermal phase fluctuations [Eq.~(\ref{TP})] and condensation ratio [Eq.~(\ref{GE})]  using a numerical iteration method. Following standard practice, we introduce the cutoff energy ${\pm}\omega_D$ for the integral in the CDW gap equation:
\begin{equation}
\frac{1}{\lambda}=-\int^{\omega_D}_{-\omega_D}\frac{f\big(E_k\!+\!v_Fp_s\big)\!-\!f\big(-E_k\!+\!v_Fp_s\big)}{2E_k}{d\varepsilon_k},  
\end{equation}
and in the condensation ratio $f_s$ [Eq.~(\ref{fs})], and introduce the cutoff ${\pm}q_{c}$ for the integrals involving bosonic excitation in the thermal phase fluctuations:
\begin{equation}
\langle{p_s^2}\rangle=\int^{q_c}_{-q_c}\frac{dq}{2\pi}\frac{q^2n_B\big(\Omega_{{\rm P}}(q)\big)}{D_0^*\Omega_{{\rm P}}(q)}.  
\end{equation}
In our specific simulations, we adopt material parameters corresponding to quasi-1D CDW material (TaSe$_4$)$_2$I, and the used specific parameter values are listed in Table~\ref{TASEI}.

\section{Discussion of critical point}

In this section, we discuss the critical points of the transition.  Firstly, the CDW depinning {crossover} temperature  $T_d$ satisfies the condition $\xi^2p_s^2(T=T_d)\gg1$, expressed as: 
\begin{eqnarray}
\xi^2\!\!\int\frac{dq}{2\pi}\frac{q^2n_B\big({\hbar}qv_F\sqrt{f_sm/m^*}\big)}{D^*_0{\hbar}qv_F\sqrt{f_sm/m^*}}\Big|_{T=T_d}\gg1\Rightarrow\xi^2\!\!\int\frac{dq}{2\pi}\frac{q^2k_BT_dm/m^*}{D_0{\hbar}^2q^2v^2_Ff_sm/m^*}\approx\frac{\xi^2k_BT_d{\pi}q_{c}}{{\hbar}v_Ff_s}\gg1.
\end{eqnarray}
 The depinning {crossover} is clearly independent of $m_{\rm P}(T=0)$ and hence is sample-independent, while the phason mass arises from the CDW pinning by impurities, commensurability effect or lattice imperfections. 

For the CDW gap transition at $T_c$, as demonstrated in the main text, this transition to normal state occurs around the temperature where the strength of the Doppler shift $|p_s(T)v_F|$ starts to exceed the CDW gap $|\Delta_0(T)|$. As mentioned in the main text, this first-order  phase transition of CDWs, which closely resembles the first-order  phase transition of the phase-fluctuating superconductivity~\cite{yang2021theory,emery1995importance,benfatto2001phase,pracht2016enhanced,fisher1990presence,fisher1990quantum,curty2003thermodynamics,li2021superconductor,dubi2007nature,sacepe2020quantum,sacepe2008disorder}, is analogous to the phase transition into the Fulde-Ferrell-Larkin-Ovchinnikov (FFLO) state in superconductors~\cite{fulde1964superconductivity,larkin1965zh,yang2018fulde,yang2018gauge,yang2021theory}, and also aligns with Landau's idea of the superfluid velocity threshold in liquid helium-4~\cite{landau1941theory,abrikosov2012methods}.  In both cases, superfluid flow abruptly ceases when the Doppler shift exceeds the energy gap of quasiparticles. To gain deep insight into the critical behavior in CDWs, we focus on the CDW anomalous correlation $\langle{\mathcal{R}_{k\sigma}}\mathcal{L}^*_{k\sigma}\rangle$, which directly reflects the existence of the CDW condensation as the characteristic quantity~\cite{hayashi2000ginzburg}. This quantity is derived as
\begin{equation}\label{annEq}
F_{k}(p_s,T)=\langle{\mathcal{R}_{k\sigma}}\mathcal{L}^*_{k\sigma}\rangle=-\frac{1}{2}\sum_{ip_n}{\rm Tr}[G(ip_n,k)\tau_1]=-|\Delta_0|\frac{f(E_{k}^+)-f(E_{k}^-)}{2E_{k}}.  
\end{equation}
From Eq.~(\ref{annEq}), the CDW anomalous correlation vanishes in the $k$ regions with $|v_Fp_s|>E_k$ [i.e., in regions with positive $v_Fp_s>E_{k}$, where the quasiparticle energy $E_{k}^{\pm}=v_Fp_s\pm{E_{k}}>0$ and hence $f(E_{k}^{\pm})=1$, and in the regions with negative $v_Fp_s<-E_{k}$, where the quasiparticle energy $E_{k}^{\pm}=v_Fp_s\pm{E_{k}}<0$ and hence $f(E_{k}^{\pm})=0$].  These regions with vanishing anomalous correlation are referred to as uncondensed regions, in which carriers no longer participate in the condensation and behave like the normal ones. The emergence of these normal-state carriers requires $|p_s|>|\Delta_0|/v_F$, leading to a threshold for the thermal phase fluctuations to destroy the CDW state. Consequently, in the self-consistent calculation of thermodynamic theory of 1D CDWs, when the strength of the Doppler shift $|p_s(T)v_F|$ exceeds $|\Delta(T)|$, the induced condensation breaking suppresses the CDW gap and hence condensation ratio $f_s$. This further enhances the thermal fluctuations to break the condensation and finally leads to the breakdown (first-order transition) of the CDW state after the  self-consistent formulation.

{\bf Physical picture.---}The discontinuous transition here is a hallmark of low-dimensional systems. It is  not of the conventional Landau mean-field type associated with a double-well free-energy landscape. This discontinuity arises from a self-consistent feedback of strong phase fluctuations, which drive an abrupt collapse of the order parameter at $T_c$.  Specifically, in the bare mean-field background, the CDW gap would vanish continuously in a second-order transition at $T_c^{MF}$. Our manuscript proposed that significant fluctuations emerge above $T_d$ due to the phason softening (i.e., CDW depinning), and these fluctuations advance the transition by causing the gap to collapse at $T_c\ll{T_{c}^{MF}}$. This scenario is, in fact, consistent with the recent experimental work~\cite{PhysRevLett.132.266504} which emphasized that the CDW near $T_c$ is fluctuation-dominated, reported a crossover from mean-field to fluctuation-dominated behavior, realized the role of the fluctuations in controlling the phase transition, although the reported crossover temperature ($\sim$225~K) in this experiment is higher than our $T_d$ and previous experimental observations.  We suspect that this difference arises from limitations of noise spectroscopy/analysis (which is an indirect probe), while noting that the underlying physics remains the same. From a more fundamental quantum-statistical perspective, in quasi-1D systems,  due to the divergence of Nambu–Goldstone-mode thermal excitations~\cite{nambu1960quasi,goldstone1961field,goldstone1962broken,nambu2009nobel,littlewood1981gauge}, the well-known Mermin–Wagner theorem forbids the establishment of true long-range order (with continuous symmetry breaking) at any finite temperatures~\cite{hohenberg1967existence,mermin1966absence,coleman1973there}. In this limit, the transition at $T=0^+$ must be  sharp.   
  In real 1D CDWs, however, impurities, commensurability effects, or lattice imperfections pin the CDW phase, thereby opening a finite excitation gap in the phason (Nambu–Goldstone) mode and suppressing divergent fluctuations. This pinning effectively circumvents the Mermin–Wagner restriction, allowing CDW order to persist at finite temperatures.  However,  as the temperature increases, the pinning potential weakens (i.e., the phason softens) and can approach a minimal value, at which point strong fluctuations  reemerge  and  are expected to drive the collapse of the CDW order that would otherwise occur only at $T=0^+$. Thus, the mechanism here remains fundamentally consistent with the Mermin–Wagner framework: CDW pinning shifts an otherwise abrupt transition at $T=0^+$ to a finite temperature $T_c$ where the pinning vanishes.

  {{In particular, we emphasize that this scenario is relevant when the depinning crossover takes place below $T_c$, thereby entering a fluctuation-dominated regime in which strong bosonic phase fluctuations play a central role in shaping the transition. This effect is expected to be especially pronounced in lattice-driven low-dimensional CDW systems, which are subject to the constraints imposed by the Mermin--Wagner theorem and exhibit a significant suppression of the phase-mode energy spectrum due to effective mass renormalization, characterized by the well-established factor $m/m^*$, thereby strongly enhancing fluctuation effects.}}

  \begin{center}
\begin{figure}[htb]
  {\includegraphics[width=6.5cm]{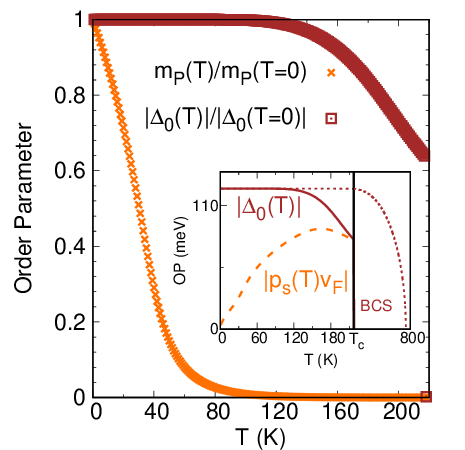}}
  \caption{{{Theoretically calculated CDW gap and phason mass of o-TaS$_3$ as a function of temperature. The inset shows the CDW gap $|\Delta(T)|$ (solid curve) and Doppler shift $|p_s(T)v_F|$ (dashed curve) from our theory and the BCS-type gap (dotted curve) versus temperature. Model parameters are fixed as follows. The coupling constant $\lambda$ is determined by fitting the experimental low-temperature CDW gap, $|\Delta_0(T\sim0)|=125~$meV~\cite{NASRETDINOVA2015180,nasretdinova2009electric}. For consistency, we take the Debye cutoff as  
  $\omega_D=|\Delta_0(T=0)|$, and the cutoff  $q_c$ in treating thermal phase fluctuations is determined by $(2q_c)^{-1}=d_{110}\approx1.47~$nm, corresponding to the interplane spacing along the [110] direction.  Other parameters of o-TaS$_3$ are not yet fully available in the literature, we adopt values from the closely related compound (TaSe$_4$)$_2$I [including $v_F=6.59\times10^7~$cm/s and $m_P(T=0)=0.9\hbar{q_cv_F}$], reflecting their structural and electronic similarity. }}}    
\label{Sfigyc}
\end{figure}
\end{center}

\section{Potential applications to other CDW materials}

The theory of the present study is developed from the basic microscopic model within a fundamental quantum-statistic approach. It, in principle, is applicable to a broad class of lattice‐driven CDW materials, although material-specific applications may require system-dependent extensions,  
depending on material-specific aspects such as dimensionality, multiband character, or more detailed ground-state properties. In the present work, to reveal the essential concepts, we have chosen a simple and well-characterized prototype, quasi-1D CDW
material  (TaSe$_4$)$_2$I. 

Here we also apply the theory to another quasi-1D CDW material, orthorhombic TaS$_3$ (o-TaS$_3$), which is a known quasi-1D CDW material~\cite{monceau2012electronic}. Similar to (TaSe$_4$)$_2$I, o-TaS$_3$ also exhibits an exceptionally well-defined quasi-1D character, a simple single-band Fermi surface, and it undergoes a Peierls transition into a fully gapped CDW state at 213~K, accompanied by a dramatic increase in resistivity. Using the ground-state parameters, our theory, as shown in Fig.~\ref{Sfigyc}, predicts a depinning {crossover} at $T_d\approx100~$K, a first-order CDW phase transition at $218~$K, and a large gap ratio $2|\Delta(0)|/(k_BT_c)\approx13.33$. These values agree with experimental data, which report $T_c=213~$K and $2|\Delta(0)|/(k_BT_c)\approx13.95$~\cite{monceau2012electronic}, while   no anomalies have yet been experimentally reported near 100~K, likely because few measurements have targeted this temperature range in o-TaS$_3$, in contrast to the extensive studies performed on (TaSe$_4$)$_2$I.  This thus serves as a concrete theoretical prediction that can be tested in future experiments.

Other quasi-1D CDW materials, such as  (NbSe$_4$)$_3$I, (NbSe$_4$)$_{10}$I$_3$, and related compounds, often involve multiband effects, partial gapping of the Fermi surface, and multiple CDW transitions~\cite{gruner1988dynamics,monceau2012electronic}. Extending our framework to these systems requires  including the multibands, analogous to the application of BCS superconductivity theory to multiband systems such as the two-band superconductor MgB$_2$. 

\section{Additional discussions}

{\sl Material parameters.---} It should be emphasized that our approach is a quantum‐statistical, first‐principles–inspired framework, so it does not involve any first-principles calculation of the material parameters (the ground‐state parameters itself), and still requires first‐principles DFT calculations or low‐temperature-limit experiments to provide these parameters as input.

{{{\sl Coulomb effects.---}The present study neglects the effect of long-range Coulomb interactions. In principle, because the charge conservation link phase dynamics to charge fluctuations, the phase gap due to Coulomb is proportional to the density of freely moving charged carriers, exactly as in plasma oscillations. Therefore, at temperatures below $T_c$, since the fermionic quasiparticle density is strongly suppressed (as realized by transport where the resistivity shows a sharp upturn upon cooling below $T_c$), the Coulomb-induced phase gap in fully gapped CDWs should be minimal. This point is  supported by previous microscopic derivations~\cite{hayashi1996topological,hayashi2000ginzburg,PhysRevB.48.1368} and very recent studies~\cite{PhysRevB.108.045148}. It is also consistent with the early experiments which have directly measured the phason gap of (TaSe$_4$)$_2$I to be only $0.03~$THz ($\sim0.12$~meV) at 6~K, measured in Ref.~\cite{DONOVAN1990721} and also mentioned in Ref.~\cite{kim2023observation}.  Based on these observations, Coulomb screening is not expected to be a dominant effect for pinning and the Coulomb-induced contribution should be insufficient to influence either the thermodynamic properties or the transition temperature $T_c$. This is consistent with the conventional understanding in both experimental and theoretical literature that these properties are governed by the CDW gap formation and by the thermal phase fluctuations.}}

{\sl Phonon-phonon interactions.---}In our model, the phason softening is described by $m^2_{P}(T)=m^2_P(T=0)\exp(-\xi^2p_s^2)$, following the self-consistent-field approximation~\cite{maki1986thermal,rice1981impurity}. This produces an ideal exponential softening. In a real material, however, additional interactions and scattering processes from other phonon branches might disrupt this behavior once the phason mass becomes sufficiently small, since such low-energy modes are inherently more susceptible to scattering and dynamical instability. A controlled, microscopic incorporation of these effects remains open, as it would require evaluating the phason self‐energy renormalization induced by interactions with other phonon branches, yielding a renormalized mass. For example, the phason Green function would be written as 
\begin{equation}
D_{P}^{-1}(\omega,q,T)=m^2_P(T)+v_P^2q^2-\omega^2-\Sigma_{P}(\omega,q,T) \Rightarrow {\bar m}_P^2(T)+v_P^2q^2-{\bar \omega}^2, 
\end{equation}
where $\Sigma_P(\omega,q,T)$ is the phason self‐energy due to interactions with other phonon branches, and $\bar{m}_P(T)$, $\bar{\omega}$ denote the renormalized quantities. Then, given a specific form of the interaction self-energy (e.g., phason–phonon interactions via three-/four-phonon vertices), one must derive the renormalization equations for $\bar{m}_P(T)$ and $\bar{\omega}$. Technically, this entails computing higher‐order Feynman diagrams with multiple scattering channels.  A rigorous calculation of this would also require advanced first-principles methods that incorporate all relevant phonon-phonon interactions. At the current stage, calculation of these potential corrections for {\sl corrective modifications} would introduce multiple phenomenological parameters. In order to emphasize that
the phason softening is the primary relaxation channel for the amplitudon, this extension is not pursued in the present study.

{{\sl Brazovskii transition.---}
As mentioned in the main text, in low-dimensional systems fluctuation effects are essential and can modify the nature of the transition away from the mean-field second-order expectation, in some cases yielding weakly first-order behavior~\cite{PhysRevB.87.134407,PhysRevLett.69.1085,PhysRevB.107.075130,PhysRevB.96.094419}. Such fluctuation-induced first-order transitions have been widely discussed in a broad range of contexts, including superconductors, liquid crystals, and helimagnets~\cite{PhysRevB.87.134407,PhysRevLett.69.1085,PhysRevB.107.075130,PhysRevB.96.094419}. While both scenarios involve fluctuation-induced first-order transitions in finite-$q$ ordering systems, the Brazovskii transition~\cite{brazovskii1975phase,PhysRevB.87.134407} in this context arises from the large phase space (large occupation) of soft modes distributed on a certain momentum shell, whereas in the present work the first-order transition is driven by self-consistent feedback of depinning-enhanced phase fluctuations into the gap equation. These two scenarios therefore correspond to distinct physical mechanisms.}
